\documentclass[sigconf]{acmart}

\AtBeginDocument{%
  \providecommand\BibTeX{{%
    \normalfont B\kern-0.5em{\scshape i\kern-0.25em b}\kern-0.8em\TeX}}}

\usepackage[linesnumbered,lined,vlined,ruled,commentsnumbered,noend]{algorithm2e}

\usepackage{booktabs}
\usepackage{times}
\usepackage{multicol}

\usepackage{tabularx}
\usepackage{graphicx}
\usepackage{longtable}
\usepackage{csquotes}
\usepackage{amsmath}
\usepackage{amssymb}
\usepackage{breakurl}
\usepackage{opera}
\usepackage{todonotes}
\usepackage{amsmath}
\usepackage{afterpage}
\usepackage{enumitem}
\usepackage{array}
\usepackage[tight]{subfigure}

\setlist{nosep}
\setlist[itemize]{leftmargin=*}
\usepackage[htt]{hyphenat}

\usepackage{titlesec}
\titlespacing{\section}{0pt}{0.1cm}{0.1cm}
\titlespacing{\subsection}{0pt}{0.1cm}{0.1cm}
\titlespacing{\subsubsection}{0pt}{0.1cm}{0.1cm}

\newcolumntype{P}[1]{>{\centering\arraybackslash}p{#1}}

\definecolor{codegreen}{rgb}{0,0.6,0}
\definecolor{codegray}{rgb}{0.5,0.5,0.5}
\definecolor{codepurple}{rgb}{0.58,0,0.82}
\definecolor{backcolour}{rgb}{0.95,0.95,0.92}
\lstdefinestyle{mystyle}{
   numberstyle=\tiny\color{codegray},
   basicstyle=\footnotesize\ttfamily,
   breakatwhitespace=false,
   breaklines=true,
   captionpos=b,
   keepspaces=true,
   numbers=left,
   numbersep=5pt,
   showspaces=false,
   showstringspaces=false,
   showtabs=false,
   tabsize=2
} 
 
\setcopyright{none} 
\renewcommand\footnotetextcopyrightpermission[1]{}

\begin{document}

\title{Tuneful: An Online Significance-Aware Configuration Tuner for Big Data Analytics}

\author{Ayat Fekry}
\affiliation{%
  \institution{Computer Laboratory\\ University of Cambridge}}
  
\author{Lucian Carata}
\affiliation{%
  \institution{Computer Laboratory\\ University of Cambridge}}

\author{Thomas Pasquier}
\affiliation{%
  \institution{Department of Computer Science\\ University of Bristol}}
  \author{Andrew Rice}
\affiliation{%
  \institution{Computer Laboratory\\ University of Cambridge}}
  \author{Andy Hopper}
\affiliation{%
  \institution{Computer Laboratory\\ University of Cambridge}}

\renewcommand{\shortauthors}{Fekry et al.}

\begin{abstract}
	Distributed analytics engines such as Spark are a common choice for processing extremely large datasets.
However, finding good configurations for these systems remains challenging, with each workload potentially requiring a different setup to run optimally. Using suboptimal configurations incurs significant extra runtime costs.

We propose Tuneful, an approach that efficiently tunes the configuration of in-memory cluster computing systems.
Tuneful combines \emph{incremental} Sensitivity Analysis and Bayesian optimization to identify near optimal configurations from a high-dimensional search space, using a small number of executions. This setup allows the tuning to be done online, without any previous training.
Our experimental results show that Tuneful reduces the search time for finding close-to-optimal configurations by 62\% (at the median) when compared to existing state-of-the-art techniques. This means that the amortization of the tuning cost happens significantly faster, enabling practical tuning for new classes of workloads. 

\end{abstract}

\maketitle
\pagestyle{plain}   

\section{Introduction}
\label{sec:introduction}
The success of big data solutions heavily relies on the prompt extraction of valuable insights from data.
The processing speed in big data systems must keep up with the ever growing data volume~\cite{idc} in a cost-efficient manner, by taking advantage of the available optimization opportunities.
The need to analyse large datasets has led to the wide adoption of Data Intensive Scalable Computing (DISC) Systems such as  Hadoop~\cite{hadoop}, Spark~\cite{spark} and Flink~\cite{flink}.
These DISC systems enable the manipulation and analysis of large amounts of data by distributing work over a cluster of machines.
They are used to help organizations make better and faster decisions, as well as in research areas such as life-sciences~\cite{sparkInBio} or physics~\cite{jacobs2016}.

One of the challenges in setting up DISC systems is to identify the right configuration in order to accelerate workload execution.
Misconfiguration can lead to either resource contention/exhaustion or under-utilization, with the former potentially triggering late runtime errors (hours after the start of a task's execution).
Consequently, developers spend significant time and resources identifying the appropriate configuration for their workload.
This has motivated work towards the automation of configuration tuning~\cite {starfish, gunther, mronline, arome, cherrypick, paris}.
The recent work on automatic configuration tuning is based on developing strategies for the exploration of the configuration search space, building on techniques such as hill climbing~\cite{mronline} and genetic algorithms~\cite{gunther}.

Generally, such explorations follow one of two strategies: In the first, a model is built to predict the execution time given a set of workload characteristics (resource consumption metrics) and a particular configuration~\cite{starfish, arome}. The model is pre-trained on numerous executions of different workload types and configurations, after which it can cheaply estimate run times for a new (workload, configuration) pair. However, tuning results are highly dependent on the model accuracy and on the similarity between workloads seen during training and actual workloads.

In the second strategy, an incremental search of the configuration space is done, feeding back information about the workload actual execution cost for a given configuration and selecting a better configuration for the next run~\cite{gunther}. Current approaches require a long search phase (around 500 executions)~\cite{bestconfig}. The practicality of such solutions hinges on the challenge of amortizing the cost of the optimisation, a function of both how fast the algorithm converges to very good configurations and how many bad (slow) configurations are explored in order to get there. 

None of the existing solutions are practical for workloads that change over time and might require re-tuning (\eg due to change in the environment or growth in the volume of data the workload receives). This is because getting optimal configurations through re-tuning requires an investment in time and resources comparable to the initial tuning costs. 
Tuneful, the system presented in this paper, makes tuning in those scenarios possible. It provides better accuracy than strategies based on pre-built models and reaches close-to-optimal configurations significantly faster than existing incremental search strategies. It does this by running the actual workloads and making better decisions about which configuration to explore next. This is done online, in the context of workloads that are run periodically. Each execution is used to pick a configuration that has the maximum probability to minimize a cost function (e.g. runtime or actual costs) next time that workload is run.
It is designed from the start to perform \emph{incremental} optimizations and make re-tuning cost-effective, requiring 62\% less search time at the median and 97\% less in the best-case, while finding configurations with similar runtimes as current state-of-the-art approaches. 
Tuneful achieves this by leveraging Sensitivity Analysis (SA) to identify the influential configuration parameters in a high-dimensional space and Bayesian Optimization (BO) using Gaussian Processes (GP) for an efficient tuning of those parameters.
\noindent The key contributions of this work are:
\begin{itemize}
	\item Finding \emph{workload-specific} influential parameters incrementally using a small number of workload executions, within a high dimensional configuration space.
	\item Applying BO to tune the influential configuration parameters incrementally using a small number of workload executions. Tuneful is the first work to propose a \emph{data-efficient}~\cite{deisenroth2015gaussian} tuning of high-dimensionality configurations.
	%
	\item Obtaining comparable or better configurations than prior work but converging to those significantly faster.
\end{itemize}


\section{Motivation}
\label{sec:motivation}

\begin{figure} [t]
	\centering
	\hspace*{0.5cm}
	\includegraphics[width=\columnwidth]{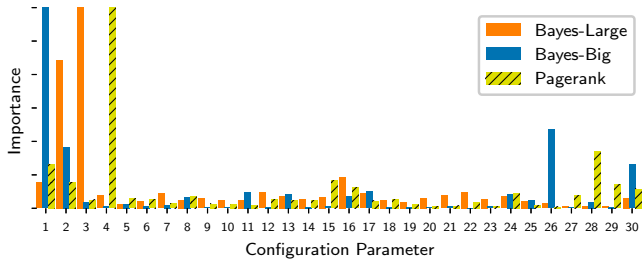}
	\vspace*{-5pt}
	\caption{Relative configuration parameter importance for 3 Hibench workloads, showing that the parameters which affect runtime the most are workload-specific.}
	\label{fig2}
	\vspace*{-7pt}
\end{figure}

\subsection{Auto-tuning is crucial in DISC systems}


\noindgras{No single configuration fits all workloads:}
Each workload has particular characteristics in terms of resource usage,
(\eg some workloads are CPU intensive, while others are memory or shuffle heavy).
Therefore there is no one single optimal configuration, but instead a workload-specific optimal configuration that is a function of its characteristics, including the distribution and size of input data.

\noindgras{Frequent workload changes:}
A change in a workload takes place either through modifying the execution logic
or through a significant change of the processed data size (\eg new records need analysis).
While the former is rare in a deployed system, the later is common in the big data world.
For example, consider a Pagerank workload, initially processing a dataset of 5 million pages. Running daily and receiving roughly 100K pages of additional data at each run, the dataset grows to 14m pages in 3 months (90 executions of the workload). Such changes can lead to significant changes in the workload's behaviour under particular configurations.

The configuration should therefore be re-tuned accordingly. While the decision about \emph{when} to re-tune is independent from the tuning strategy, its \emph{feasibility} relies on how expensive is to re-run tuning in the setting of gradually changing workloads. 

Existing approaches require a significant number of executions for finding close-to-optimal solutions. In particular, the number of executions required generally exceeds the number of times the workload might run before re-tuning is necessary, making such solutions impractical.
For example, the \texttt{BestConfig}~\cite{bestconfig} system requires hundreds of samples to identify a good configuration, exceeding the 90 ``normal'' runs of our Pagerank example workload over the 3 months period. We will focus on developing strategies that are able to identify good configurations from a significantly smaller number of samples.

\noindgras{Significant number of execution samples}:
The search space for good configurations grows exponentially with the number of configuration parameters.
Spark~\cite{spark} has over 180 configuration parameters. Not all of them impact performance, but some of the relevant ones have large intervals of possible values.

For example, the search space of three Spark configuration parameters (\ie memory per executor, cores per executer and spark memory fraction) contains more than 4000 combinations, assuming a cluster of nodes with 60 GB memory and 16 processing cores.
In practice, tuning the configuration using evolutionary or hill climbing approaches~\cite{gunther, opentuner} requires hundreds of execution samples, representing thousands of hours of computational power and associated monetary cost.
It is therefore crucial to minimize the number of configurations evaluated to reach an optimal configuration.
Bayesian Optimisation (BO) strategies have been applied recently as a solution to this problem.  However, BO  fails to provide quick convergence in high dimensional configuration spaces (more than 10 parameters)~\cite{shahriari2015taking}.

\subsection{The significance of configuration parameters case study}
To enable quicker convergence of BO in a high dimensional space,
some DBMS work proposed to reduce the dimensionality of the configuration parameters, building on techniques such as factor analysis~\cite{ottertune} and sensitivity analysis~\cite{rafiki}. 
This work runs intensive offline benchmarks to find a set of system-wide influential parameters that impact the performance of all the workloads. 
However, this approach is hard to adopt for DISC systems due to the high diversity of the running workloads (e.g graph analytics, machine learning, SQL, and text analysis). This diversity poses the need for a dynamic understanding of each workload in terms of its influential parameters and should perform their tuning accordingly.

In this case study, we show that the configuration parameters that have the biggest impact on the runtime vary depending on the workload characteristics.
We experimented with three workloads from Hibench~\cite{hibench}: Pagerank and two Bayesian classifiers with different input sizes, we executed each workload 100 times using random configurations for the 30 parameters in~\cite{tuneful_link}(we selected those parameters as they represent a superset of the ones used in the related work~\cite{bestconfig, yu2018datasize}).
~\autoref{fig2} depicts the significant configuration parameters for the three workloads.
The calculation of the significance is based on the contribution of each configuration parameter in predicting the execution time, as determined from the 100 executions of each workload. We give more details on the importance calculation of each parameter in~\autoref{sec:approach:sig}.
As expected, for each of these workloads, the set of significant configuration parameters is different.
If we merge the influential parameters of the various prospective workloads to build a system-wide set of significant parameters, we will end up with more than 10 parameters in total and ultimately a slow convergence of the BO. This motivates exploring the influential parameters \emph{efficiently} at the workload level and not system-wide.

\section{Background}
\label{sec:background}

\begin{figure}[t]
	\centering
	\includegraphics[height=5cm]{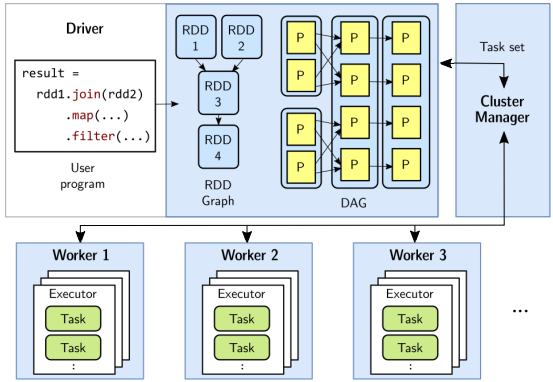}
	\vspace{2mm}
	\caption{Spark internal architecture, redrawn with further edits from ~\cite{sparkArch}}
	\label{spark}
	\vspace*{-15pt}
\end{figure}
In this section we provide an overview of Apache Spark~\cite{spark},
which is the DISC system we use as our tuning case study.
We then briefly describe BO and GP and how they optimize the exploration of the configuration space.

\subsection{Spark}

Spark is a cluster computing framework. It has been developed to overcome the limitations of the MapReduce~\cite{mapreduce} paradigm in handling iterative workloads, and is widely adopted for in-memory big data analytics.
MapReduce forces mappers to write data to disk for reducers to read,
which consumes significant I/O resources for iterative applications.
Spark keeps the data in memory as Resilient Distributed Datasets (RDDs)~\cite{rdd}, a choice that significantly reduces I/O costs and speeds up iterative job execution time by up to 10X compared to Hadoop~\cite{zaharia2010spark}.
RDDs are immutable collections distributed over a cluster of machines to form a restricted shared memory, with each RDD consisting of a set of partitions.
~\autoref{spark} shows how Spark works internally.
Users write a program and submit it to the Spark Driver, a separate process that executes user applications and schedules them into executable jobs.
The Spark programming model is based on two types of function: transformations and actions.
Transformations represent lazy computations on the RDD that create a new RDDs (\eg map, filter).
Actions trigger computation on an RDD and produce an output (\eg count, collect).
When an application invokes an action on an RDD, it triggers a Spark job.
Each job has an RDD dependency graph containing all the ancestor RDDs,
representing a logical execution plan for the set of transformations.
The RDD graph is mapped into a Directed Acyclic Graph (DAG) defining the physical execution plan: a split of the job into stages,
the dependencies between stages and the partitions processed in each stage.
The Driver uses this DAG to define the set of tasks to execute at each stage.
Typically, an RDD partition is given as input to a stage and is processed by a Spark task.
Finally, the driver sends tasks to the cluster manager, which assigns them to worker nodes.
A worker node can have multiple executors,
with each of them being a process executing an assigned task and sending the result back to the driver.

Examples of Spark configuration parameters that influence how it works internally are the number of executor instances, size of memory per executor, number of cores per executor, the size of shuffled data buffer or the size of the off-heap memory, etc.



\subsection{Bayesian Optimization and Gaussian Processes}
BO~\cite{mockus2012bayesian} is a method for minimizing blackbox functions~$f$ iteratively, using a limited number of samples.
This is useful when it is expensive to evaluate $f$ at a given point (such as running a big-data workload with a given configuration).
BO is characterized by its \emph{prior model} and \emph{acquisition function}: the prior model represents a space of possible target functions $f$, and the acquisition function guides the selection of the next evaluation point based on the prior modelled knowledge.
One of the widely accepted prior models for BO is Gaussian Process (GP).
It represents a distribution over functions (a sample drawn from this process is a function) with given mean and covariance. Here, the mean function describes expected values at each point and the covariance function defines the smoothness of the functions which can be drawn as samples, encoding prior assumptions about the data that we want to model~\cite{rasmussen2004gaussian}.

The GP maintains a probabilistic belief about what functions $f$ are possible, given known characteristics and already seen data. This belief is updated by using an acquisition function, which determines the best point sample of $f$ to take next. After sampling, the prior belief about possible functions $f$ is updated and a new sampling decision can be made, iteratively. At each step, the posterior distribution has filtered-out functions not consistent with the sampled data and will ideally have a narrower candidate function space.

The GP acquisition function represents the metric by which the GP picks the next input sample to improve the probabilistic model function. It is typically a function that is cheap to evaluate at a given point $x$ and its value is proportional to how useful evaluating $f(x)$ would be for the optimisation problem.
Various acquisition functions have been proposed to define the way the GP samples the input space, \eg random, sequential, Probability of Improvement (PI) or Expected Improvement (EI)~\cite{villemonteix2009informational, srinivas2009gaussian, hoffman2014correlation}. We discuss GP as it applies to Tuneful in~\autoref{sec:approach}.

\section{Approach}
\label{sec:approach}

To reduce the cost and make incremental tuning a viable proposition, Tuneful leverages GP to efficiently find near-optimal configurations using a small number of samples.
GP uses metrics captured from each execution in order to update its model about how a workload behaves and then picks the next best configuration to explore.
The strategy has been shown to be a data-efficient learning approach in a different domain (robotics control)~\cite{deisenroth2015gaussian}.
However, directly applying GP to the high-dimensional configuration parameter space of Spark is insufficient for obtaining satisfactory results,
because in practice GP converges relatively slowly when tasked with approximating functions in a high-dimensional space~\cite{tripathy2016gaussian}. 
This can be explained as an effect of GP's reliance on the Euclidean distance to define input space correlations.
Euclidean distance becomes less informative as the input space dimensionality increases~\cite{bengio2006curse} and the number of samples required to learn the model grows exponentially (the curse of dimensionality~\cite{bellman1956dynamic}).
To address this shortcoming, Tuneful first determines the configuration parameters that influence the workload execution time the most\footnote{We focus on execution time through the paper, but the algorithm works similarly for any defined cost function.}, and then builds a GP cost model in order to tune only those. However, finding influential parameters typically requires lots of executions in itself. 

Tuneful does not require expensive offline phases for significant parameters identification or tuning.
It is designed to provide online \emph{incremental} optimization, first suggesting configurations that enable a quicker exploration of the influential parameters, and then offering subsequent configurations that optimize those influential parameters based on the prior runs. 
Ultimately, we show that Tuneful enables \emph{data-efficient} tuning in a high-dimensional space with a minimal overhead.

\subsection {Identifying Significant Parameters}
\label{sec:approach:sig}

Virtually all complex systems have numerous configuration parameters that can be set to user-given values. However, only a small subset of those parameters has a significant impact on workload overall performance.
\autoref{fig2} supports this conclusion, showing that out of the selected 30 Spark configuration parameters only a small fraction influence the execution time significantly. Tuning the others (non-influential parameters) simply wastes resources and does not bring the algorithm closer to the optimal solution\footnote{In our experiments we generally observe that the importance of a given parameter is independent of others: we are not claiming that the parameters themselves are independent, just that the choice of a value for a given parameter tends not to affect the relative importance of the other parameters.}.


Tuneful does not define system-wide influential parameters, since this approach is expensive and hard to apply for a DISC system receiving diverse workloads.
Tuneful identifies workload-specific influential parameters.
For each workload, we \emph{incrementally} define the set of parameters that impact its performance the most.
We developed a method of Sensitivity Analysis (SA)~\cite{borgonovo2016sensitivity} that is efficient in our problem domain. In our context, SA studies how the variation of the cost function (execution time, considered as output) can be attributed to the different configuration parameters (inputs).
There are two main types of SA:
\begin{itemize}
\item Quantitative: methods that assign a numerical value to the influence of each input parameter on the total variance in output (execution time). 
The Sobol method ~\cite{sobol2001global} is a representative example, requiring hundreds of executions to determine influence metrics accurately.
%
%
\item Qualitative: methods that only aim to distinguish between influential and non-influential input parameters, requiring fewer executions than quantitative SA methods.
They compute an ordering amongst available parameters, starting from the most influential.

Screening~\cite{borgonovo2016sensitivity} is one of the most common approaches to qualitative SA, based on sampling the input space to rule out non-influential parameters. One-at-a-time (OAT)~\cite {borgonovo2016sensitivity} is a type of screening that we build upon, varying one parameter at a time and fixing the remaining inputs to measure the influence of the non-fixed input on performance.
Tuneful uses a similar idea as part of a hybrid SA approach. Standalone, OAT requires a number of executions in the order of $d$, the number of dimensions of the input.
\end{itemize}

A second idea to reduce exploration costs that Tuneful adapts to meet its goals is using meta-models (proposed by Steenkiste \etal~\cite{van2016sensitivity} to reduce costs for quantitative methods). In our case, the meta-model is created to estimate the execution time given a set of configuration parameters. Sensitivity measures such as Sobol indices can then be computed using the model instead of actual executions. Again, building an accurate model for our high-dimensional configuration space requires hundreds of executions.
We solve this issue by using the insight that in DISC workloads only a few parameters have a strong influence on performance, and that those \emph{dominate} regardless of the values of the other, low-influence parameters. We leverage this insight and propose a new hybrid algorithm for the identification of significant parameters.

\textbf{Key idea:} Tuneful's algorithm combines the OAT screening approach from qualitative SA with meta-modeling from quantitative SA to  \emph{incrementally} distinguish the contribution of influential parameters, using small number of workload executions.
 
\makeatletter
\def\BState{\State\hskip-\ALG@thistlm}
\makeatother
\vspace*{-5pt}
\begin{algorithm}[!ht]
	\SetAlgoLined
	\DontPrintSemicolon
	\SetKwInOut{Input}{Input}\SetKwInOut{Output}{Output}	
	\Input{$d, \alpha, n , n\_SA\_rounds, P, R$}
	\Output{$P_{s}=P_{\alpha*d}$}
	{
		$X_i= sample(P , R , P_{\mathrm{fixed}})$ \label{SAalg:line:sample}\;
		run workload using $X_i$ and get $C_i(X_i)$\;
		$n\_executions \gets n\_executions+1$\;
		\If{$n\_executions > n$ and $n\_SA\_rounds > 0$}{
		build  $M (\mathbf{X}, \mathbf{C})$ \label{SAalg:line:model}\;
		find\_the\_importance  $imp\{p_1,...,p_d\}$  using $M$\label{SAalg:line:gini}\;
		find $P_{\alpha*d} \subset P$ with the highest importance\label{SAalg:line:find_imp}\;
		$P_{\mathrm{fixed}} \gets P - P_{\alpha*d}$ \label{SAalg:line:fixate}\;
		$d \gets \alpha*d$\;
		$n\_SA\_rounds \gets n\_SA\_rounds-1$\;
        $n\_executions \gets 0$;
        
	}	
	}
\caption{Significant Parameter exploration}
\label{alg1}
\end{algorithm}
\vspace*{-8pt} 
 
We describe the algorithm in the rest of this subsection. It is repeatedly called during the SA phase for each execution of a workload. Its input arguments are as follows: 

\begin{itemize}
\item $d$ the number of configuration parameters considered;
	\item $P=\{p_1, ...,p_d\}$ the configuration parameters;
	\item $R=\{r_1,...,r_d\}$ the range of values of each $p_i$ configuration parameter;
	\item $\alpha$ the fraction of configuration parameters retained in each SA round;
	\item $n$ the number of samples required per SA round;
\end{itemize}

Conceptually, we consider three global variables: $n\_SA\_rounds$, the number of SA rounds, $n\_executions$, initialized to 0, and $P_{\mathrm{fixed}}$ initialized to the empty set. Variable states are maintained between calls, so that $P_{\mathrm{fixed}}$ grows between SA rounds.

Let $X_i=\{x_{i1} ,x_{i2}, ... x_{id}\}$ be a particular configuration chosen by our algorithm, $C(X_i)$ its execution cost, and $M(\mathbf{X},\mathbf{C})$ the constructed meta-model that maps a given configuration $X$ to its execution cost $C$, used to distinguish the influential parameters.
The goal is to identify $P_{s}=\{p_1, p_2,..p_s\}$ where $|P_{s}| < |P|$ such that $P_{s}$ contains the \emph{selected} top $s$ influential parameters. 
\autoref{alg1} shows the steps of our SA algorithm:



The algorithm is run for each workload execution, until the influential parameters are determined. We perform incremental sampling at line~\ref{SAalg:line:sample} using low-discrepancy sequences~\cite{sobol1998quasi},
which represent numbers that are evenly distributed in a given space to provide a \emph{quicker} coverage. 
This quick coverage is crucial in the high-dimensional space as exploring the whole space is not a viable option.
Once we have $n$ samples and their associated $C(X_i)$,
we use them to build a Random Forest Regression (RFR) model $M$,
approximating the true dependence between $C$ and $X$ (line~\ref{SAalg:line:model}).
We use RFR as it is an ensemble learning algorithm that allows us to boost the accuracy of the model predictions while avoiding model overfitting~\cite{liaw2002classification}. It consists of multiple decision trees, each trained on a different part of the dataset. This improves the prediction accuracy compared to a single learning model.
At line~\ref{SAalg:line:gini}, the importance of each input configuration parameter in $M$ is calculated based on Gini importance~\cite{weymark1981generalized}.
This is a measure of each feature's contribution to the prediction of the execution time. 
Gini importance considers the number of times a given feature is used in a tree split, across all the forest trees.
Important features are used more frequently in decision tree splits, so they have higher Gini importance.

We then select the highest influence parameters (line~\ref{SAalg:line:find_imp}), a fraction $\alpha$ of the total $d$ parameters, and consider the remaining ones as non-influential (line~\ref{SAalg:line:fixate}).
Those are fixed for the remaining rounds to the mean of their value range and the next iteration is started to determine the most influential parameters among the ones that can still vary.
The algorithm stops after the pre-defined number of SA rounds $n\_SA\_rounds$.

\noindgras{Design choices:} 
We empirically observed the impact of different values of $\alpha$ from 0.1 till 0.9. While a very small value for $\alpha$ leads to pruning influential parameters, high values for $\alpha$ will eventually lead to a wider search space that includes many uninfluential parameters and slows the identification of the highly-influential ones.
We set $\alpha=0.6$ in the SA stage as it represents a good compromise between the accuracy of detecting the influential parameters and bounding the number of SA runs.
We set $n$ to 10 execution samples, as it guarantees generating an RFR model of an acceptable accuracy (less than 40\% error), enabling a good \emph{approximation} of the true dependence between the configuration parameters and execution time.
$n\_SA\_rounds$ needs to be selected to provide good guarantees, while minimizing cost. We discuss how we do this in practice for Tuneful in section~\ref{sec:evaluation:sig-conf}.


\subsection {Configuration Tuning}

\label{sec:approach:tuning}

To perform data-efficient tuning, we aim to get as close to the optimal configuration as possible, using the minimum number of executions.
We chose GP due to its data efficient learning performance, which makes it an effective approach for modelling expensive functions~\cite{deisenroth2015gaussian, cherrypick}.
Furthermore, GP is non-parametric, which means that it does not need users to pre-commit to the shape of the function that models the cost.
This flexibility allows GP to model the runtime of heterogeneous workloads and the influence of various configuration parameters on them.

\noindgras{Problem formulation:}
We define the objective function that the GP tries to minimize as the execution time of the workload.
Tuneful's proposed approach is extensible and can handle any other objectives such as minimizing energy consumption, cluster utilization or a weighted sum of objectives.

\noindgras{Acquisition function:}
We use the GP Expected Improvement (EI) with Markov Chain Monte Carlo (MCMC) hyperparameter marginalization algorithm~\cite{spearmint} as our acquisition technique.
This is an EI based~\cite{brochu2010tutorial} function, which iteratively selects the next configuration sample as one that has the highest potential to minimize the objective function.
It has the distinctive advantage of not requiring any external tuning of GP hyperparameters unlike other acquisition functions such as GP upper confidence bound (GP-UCB)~\cite{rasmussen2004gaussian}.
Other acquisition functions are possible, such as random, sequential and Probability of improvement (PI). We use EI MCMC as it has shown better performance compared to other acquisition functions across a wide array of applications and test cases~\cite{spearmint}. 

\noindgras{Prior and covariance functions:}
We assume that the multi-dimensional function from a set of configuration parameter values to the runtime cost can be modelled by a Gaussian Process (this is a Gaussian Process prior model).
We made this assumption as GP can provide a wide range of flexible non-parametric statistical models over the function space.
We choose the ARD Matern 5/2 kernel~\cite{rasmussen2004gaussian} as the GP covariance function, because it is able to control how smooth the estimated function is along each of its dimensions independently. This allows Tuneful to learn the objective function quickly and just sample the data that most likely has the minimum objective function, while leaving the other less useful points unexplored. This kernel has been successfully adopted for modelling practical functions~\cite{spearmint}.

\noindgras{Starting points:}
We build the GP cost model incrementally, starting with just three samples generated using low-discrepancy sequences~\cite {sobol1998quasi}.
Then the model is improved after each further execution, picking the next candidate configuration that we estimate to reduce the execution time.

\noindgras{Stopping Criteria:}
The GP modeling stops after suggesting a minimum of n samples (e.g 10 samples) and then after the expected improvement (EI) drops below 10\%.
We made this decision to make sure that we balance between the exploration of the tuning space and exploitation of the best configuration found.

\section{Implementation}
\label{sec:implementation}
\begin{figure}[t]
	\centering
	\includegraphics[width=\columnwidth ,height=6cm]{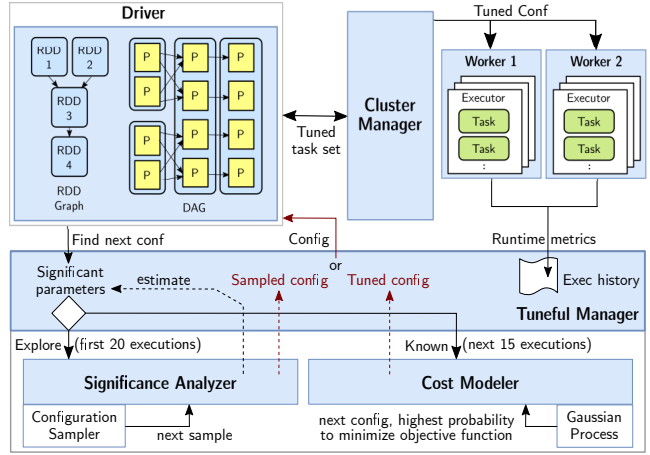}
	\vspace{1.5mm}
	\caption{Tuneful architecture and integration with Spark}
	\label{diag}
	\vspace*{-15pt}
\end{figure}


We show how Tuneful can be seamlessly integrated into Spark in order to automatically tune configurations. We have tested the integration both in AWS and Google Cloud services, without requiring changes to the Spark binary that is used.
Tuneful consists of four components: the Configuration Sampler, the Significance Analyzer, the Cost Modeller and the Tuneful Manager.
In the rest of this section, we describe Tuneful's components and their interactions with Spark, as shown in \autoref{diag}.

When a workload execution starts, Spark's driver calls into Tuneful, registered as an extension. In response, it returns the configuration that should be used next.

\textbf{Tuneful Manager:} This component first identifies the workload being submitted and retrieves its profile if it exists.
During a first phase, configurations are obtained through the Significance Analyzer aiming to identify the significant parameters (as described in \autoref{sec:approach:sig}).
In a second phase, once those parameters have been identified, the Cost Modeller takes over to minimise the objective function (as described in \autoref{sec:approach:tuning}).
Once a workload execution is finished, relevant performance metrics are provided to Tuneful and used either by the Significance Analyzer or the Cost Modeller.
This process is performed automatically every time a workload is scheduled for execution, identifying the significant parameters incrementally in the first phase then optimising the configuration by reducing cost of the objective function iteratively.
We demonstrate the practical effectiveness of the approach in \autoref{sec:evaluation}.

The Tuneful Manager monitors the performance of the workload over time to determine if reconfiguration is necessary.
As we described in \autoref{sec:motivation}, this can be due to a number of factors, such as changes to the size of the data to be analysed, or to the workload logic or changes in the underlying physical architecture of the cloud platform hosting the workload.
Once performance degradation according to some metric is detected, the Tuneful process restarts (i.e. Tuneful suggests a configuration that helps with identifying new significant parameters then tuning them, following ~\autoref{alg1} steps again). Currently, we simply define this degradation as throughput drop over time (e.g. more than 20\% drop in the amount of data processed per second over a fixed time window).
Employing more complex techniques such as workload characterization for automatically detecting the need for re-tuning is an interesting area for future work.

\textbf{Significance Analyzer:}
The Significance Analyzer finds the influential configuration parameters using ~\autoref{alg1}. It uses the Configuration Sampler to sample values for these potential influential parameters and fixes the non-influential ones. The implementation uses Python3 and the Scikit library~\cite{sci}.

\textbf{Configuration Sampler:}
During the phase that explores parameter significance,
the Configuration Sampler is used by the Significance Analyzer to sample the configuration values in a manner that accelerates the coverage of the exploration space.
It uses low-discrepancy indices, which provide a good coverage of the sampling space ~\cite{sobol1998quasi}.

\textbf{Cost Modeler:}
The Cost Modeler uses GP optimization to build the configuration cost model of a workload.
The model is built incrementally and the GP suggests the next configuration that has high potential to minimize the objective function.
To implement this module, we used Spearmint~\cite{spearmint}, which is a Python Bayesian optimization library.

\textbf{Example usage:}
In order to use Tuneful, a Spark user simply adds Tuneful library as a dependency and TunefulManager as an extra Spark listener while submitting his workload to Spark. In other words, Tuneful can run on an unmodified Spark infrastructure. ~\autoref{listing:implementation:usage} shows an example of using Tuneful when tuning the Bayes workload.

\begin{lstlisting}[float=t,language=bash, style=mystyle, caption=Tuneful example usage, label=listing:implementation:usage]
$ ./bin/spark-submit
  --jars tuneful-with-dependencies.jar
  --conf spark.extraListeners=TunefulManager
  --class BayesClassifier.java
   bayes-workload.jar
\end{lstlisting} 

\section{Evaluation}
\label{sec:evaluation}
We evaluate Tuneful in two stages: first, we examine the properties of the algorithm proposed for picking significant configuration parameters; then, we examine Tuneful as a whole performing online tuning of typical cloud-computing workloads. For the latter we consider the savings in execution time obtained through tuned configurations and the search time required to get close to the optimal when compared to three state-of-the-art approaches (Opentuner, Gunther and RandomSearch).

\subsection{Experimental setup}

\noindgras{Cluster and configuration specification:}
We use a cluster of 20 Google Compute Engine~\cite{gcp} instances (1 driver + 19 workers), with the driver being an \textit{n1-highmem-8} instance with 8 vCPUs, 52 GB memory and 300GB storage and the 19 workers being \textit{n1-standard-16} instances with 16 vCPUs, 60 GB memory and 500GB storage each.
The total cluster memory and storage size is 1.2 TB and 5.48 TB respectively. We also use a smaller cluster of 4 AWS \textit{h1.4xlarge} instances to validate the robustness of Tuneful, but experiments are run on the 20 nodes cluster unless otherwise mentioned.
We use HDFS~\cite{hdfs} version 2.7 for accessing the shared data and Spark version 2.2.1 as the system under tuning.
We tune 30 configuration parameters that cover Spark memory, processing, shuffle and network aspects,
with approximatively $2\cdot10^{41}$ configurations possible in total (this represents the size of the search space).
A list of the configuration parameters and their ranges are in~\cite{tuneful_link}]. We use the same ranges when evaluating the other tuning approaches.

\noindgras{Applications:}
We chose 4 workloads of different characteristics to experiment with the effectiveness of Tuneful in searching for close-to-optimal configurations. The workloads are chosen from the well known big data benchmarks (Hibench~\cite{hibench} and TPC-H~\cite{tpc-h}):
1) \emph{Bayes} is a workload that builds a bayesian classification model, the total executors input is 350GB.
2) \emph{Pagerank} is a graph analytics workload that ranks the influence of graph vertices, and we use the Hibench-defined \textit{huge} data size.
3) \emph{Wordcount} is a textual analysis workload that counts word occurrences in 320GB of input.
4) \emph{TPC-H} is a benchmark for big data systems that runs 22 decision support SQL queries. We use SparkSQL to run these queries on data of scale factor 20.
We chose this scale to limit the expenses of our experiments; however, we make sure that this scale is representative enough, with 300GB of total input to executors.
\begin{figure} [t]
	\centering
	\vspace*{-17pt}
	\hspace*{-10pt}
	\includegraphics[]{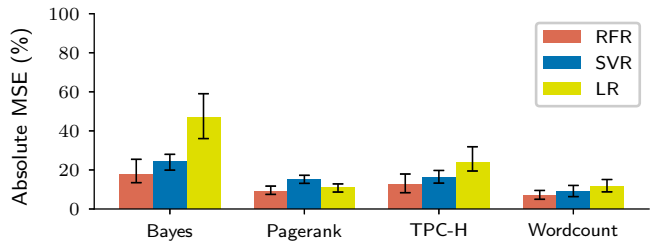}
	\caption{Error across the different models when computing influential configuration parameters using a large number of samples (lower is better).}
	\label{figure:model_accuracy}
	\vspace*{-10pt}
\end{figure}

\begin{figure} [t]
	\centering
	\vspace*{-15pt}
	\hspace*{-10pt}
	\includegraphics[]{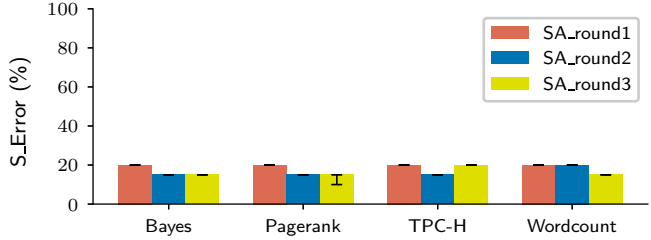}
	\caption{Weighted sensitivity error over each of three SA rounds. The error remains low across all workloads (lower is better).}
	\label{figure:SA_rounds_accuracy}
	\vspace*{-15pt}
\end{figure}

\begin{figure} [t]
	\centering
	\vspace*{-15pt}
	\hspace*{-10pt}
	\includegraphics[]{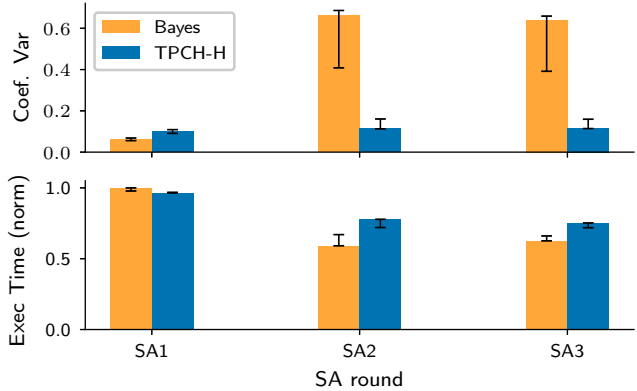}
	\caption{Coefficient of Variation (CV, $c_\mathrm{v}=\frac{\sigma}{\mu}$, top) and normalized workload execution time (bottom) when using the parameters picked after each SA round.}
	\label{figure:COV}
	\vspace*{-7pt}
\end{figure}

\subsection{Significant Parameters Exploration}
\label{sec:evaluation:sig-conf}

In this section, we evaluate the accuracy with which our algorithm detects significant configuration parameters in each SA round. We start by estimating a ground truth parameter importance for each workload, running known SA algorithms (requiring a large number of sample executions). We then compare this with the output of our algorithm (identification from small number of executions). 

\noindgras{Estimated significant configuration parameters:}
We run each workload 100 times with different configurations sampled using low-discrepancy indices, and build a RFR execution time prediction model.
From this, the most significant configuration parameters are selected using Recursive Feature Elimination (RFE)~\cite{guyon2002gene}.
To make sure that the built model has an acceptable prediction error,
we use 20\% of the samples as the test dataset and exclude them from the training data when building the model.
~\autoref{figure:model_accuracy} demonstrates the accuracy of the RFR model compared to other other strategies, namely, support vector Regression (SVR) and Linear Regression (LR). The $y$ axis shows the absolute mean square error, and the plot displays median values, with bars for the 10th and 90th percentile.

The RFR model has a median error less than 20\% across all workloads, and also maintains the lowest error for all the workloads at the 90th percentile compared to SVR and LR. We conclude that it represents the best estimate of the true significant parameters and select the top 6 of those (out of 30) for comparing with the SA stage of Tuneful (that also selects the top 6). We make this selection following the Pareto principle (80/20 rule).


%

\noindgras{Tuneful detects the significant configuration parameters within 20 samples:}
To validate our algorithm, we measure the classification error $S_{\mathrm{error}}$, representing the proportion of the best estimated significant parameters missing in the output of~\autoref{alg1} (the top 6 important parameters in $P_{s}$ after each round).
$S_{\mathrm{error}}$ is weighted based on importance,
with each configuration parameter classified either as high-influence or low-influence according to its normalised importance (as determined by RFR).
The error weight of misclassifying a configuration parameter varies depending on its class, as follows: the weight of misclassifying all the high-influence parameters is 0.8 of the total error,
while the weight of misclassifying all the low-influence ones is 0.2 of the total error.
~\autoref{figure:SA_rounds_accuracy} shows the $S_{\mathrm{error}}$ associated with each SA round, 
We run our entire experiment 10 times to plot the median, 10th percentile and 90th percentile.

Across all workloads, Tuneful reaches a 90th percentile error of  20\% or less using only 2 SA rounds, detecting all the highly-influential parameters. For TPC-H, the error slightly increases when running 3 SA rounds. This can happen with all workloads having very few influential parameters, and it means that in top 6 some non-influential parameters have been selected; those are unstable across the SA rounds and differ from the estimated ground truth. We rely on the GP phase later to detect those non-influential parameters and focus the tuning towards the high-influence ones.
We experimented tuning the configuration of TPC-H workload using the parameters detected by the second and third SA rounds and found similar results as the high-influence parameters were already detected by the second SA round.
Taking this into account, we only use two SA rounds and pick the parameters selected by the second, as a compromise between exploration (number of samples) and exploitation (information extracted from each round).

We also experimented on a smaller cluster of 4 AWS \texttt{h1.4xlarge} instances to validate that two SA rounds is enough over clusters of different resources.
~\autoref{figure:COV} (top) shows the coefficient of variation (CV, defined as the ratio between the standard deviation and the mean) in execution time for the Bayes and TPCH-H workloads if we picked parameters selected by different SA rounds for tuning. At the 90th percentile, the CV at the first SA round is small as the picked parameters for tuning do not include all the highly-influential ones. The parameters detected as important by the second SA round have higher influence, leading to higher variation (the changes made by the tuning stage to the values of parameters selected as important lead to wider variations in execution time). The third SA round has only a marginal change over the previous round.
~\autoref{figure:COV} (bottom) shows the execution time when using tuned parameters picked by each SA round. Similarly, at the 90th percentile, the parameters picked by the second SA round achieve a better execution time compared to the first SA round, with only marginal improvements made by a third SA round.

\begin{figure} [t]
	\hspace*{-5mm}
	\includegraphics[]{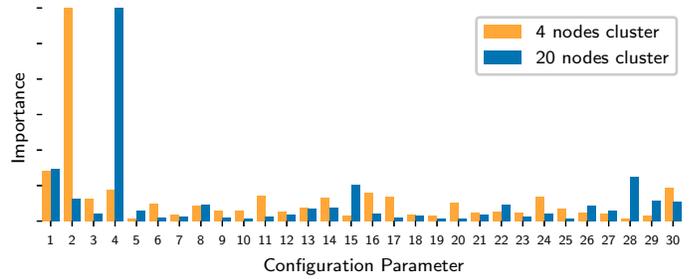}
	\vspace*{-7pt}
	\caption{Relative configuration parameter importance for the Pagerank workload executed in 2 different clusters. Parameters with the largest impact on runtime can be cluster-specific.}
	\label{diff_clusters}
	\vspace*{-15pt}
\end{figure}

\noindgras{Significance-awareness in different environments:}
~\autoref{diff_clusters} shows configuration parameter importance for the Pagerank workload across two different environments: a AWS cluster of 4 nodes and a Google compute engine cluster of 20 nodes. The calculation of significance is based on the contribution of each configuration parameter in predicting the execution time, as determined from 100 executions on each cluster.
When we deployed the Pagerank workload in the two clusters, Tuneful detected some entirely expected differences, e.g. CPU and the parallelism level (param. 4) being the most important in the large, over-provisioned cluster and memory (param. 2) the most influential in the small cluster. However, there are also less obvious differences in parameter importance: speculative execution of tasks (param 15) and whether to compress variable broadcasts (param 28) being more important on the large cluster, and the choice of the serializer (param 30) being more important in the small cluster. Overall, the algorithms we propose are able to tune accordingly for changes in the environment that reflect indirectly on how configuration values map to workload runtime.

\begin{figure} [t]
	\centering
	\vspace*{-15mm}
	\hspace*{-5mm}
	\includegraphics[]{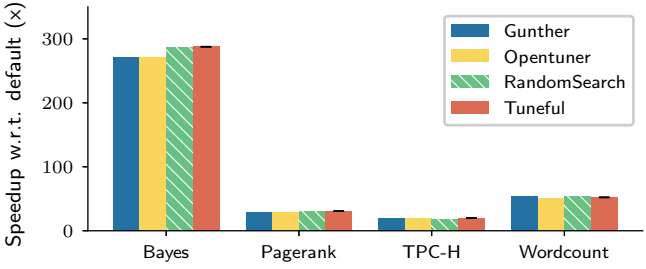}
	\caption{Execution time acceleration (X times) w.r.t Spark default configuration (higher is better).}
	\label{figure:exec_time}
	\vspace*{-15pt}
\end{figure}

\subsection{Tuning Effectiveness and Efficiency}
\label{sec:evaluation:tuning-efficiency}

We allow each state-of-the-art system a maximum budget of 100 executions for reaching a stable tuned configuration. Then we compare the configurations picked by Tuneful with the configuration tuned by:
1)\emph{OpenTuner}~\cite{opentuner}, a general tuning system that uses ensembles of search techniques such as hill climbing, differential evolution, particle swarm optimization and pattern search.
OpenTuner evaluates which techniques perform well over a window of time and picks them more frequently than the ones that have a poor performance (those can even get disabled). We selected OpenTuner as it covers a wide range of search algorithms.
2) \emph{Gunther}~\cite{gunther}, is a Hadoop configuration tuning system that leverages genetic algorithms to search for good configurations.
To compare Gunther with Tuneful, we implemented it for Spark based on the details given in the paper~\cite{gunther}, choosing a population size of 60 and evolving for 20 generations (as reasonable limits for the number of workload executions required).
3) A configuration picked through \emph{Random Search}.
Other published work either uses search techniques already covered by OpenTuner, or the implementation details were too sparse to reproduce the approach.

\noindgras{Metrics:}
We use three metrics to evaluate Tuneful:
1) \emph{Execution time saving}, the amount of execution time saved by Tuneful and competing approaches with respect to the default configuration. The target here is to obtain tuned configurations similar to what state-of-the-art tuners achieve.
2) \emph{Search Cost}, the amount of time and actual cost (in \$) required by each system to find good configurations while repeatedly running workloads in a cloud environment. The target is to get close-to-optimal configurations (within 5\% of the estimated best configuration) significantly faster than the state-of-the-art.
3) \emph{Amortization speed}, the number of needed workload executions to amortize the tuning costs. The target is to amortize the tuning cost after a small number of workload executions.

The Tuneful results are always presented as the median of 10 runs, with bars for the 10th and the 90th percentile.

\noindgras{Tuneful finds configurations comparable to the state of the art tuning systems:}
\autoref{figure:exec_time} shows that at median, Tuneful is able to obtain effective configurations for all workloads, with a very small inter-percentile (10-90) range. Tuneful maintains a comparable performance to the other non-incremental tuning systems. 
We are comparing against the default configuration as a baseline usable across workloads. The relative differences between algorithms are relevant rather than the precise acceleration figures (in a realistic setting those would be lower when starting from reasonably hand-tuned configurations).
In evaluating the execution time acceleration with the best configuration found by each tuner per workload, we don't penalize other algorithms if they are slow in finding good configurations. We therefore allow each of them to use 100 execution samples per workload (our fixed maximum budget). In comparison, we present the results of Tuneful using just 35 execution samples per workload (as it would normally be used).

\begin{figure} [t]
	\centering
	\vspace*{-15mm}
	\hspace*{-5mm}
	\includegraphics[]{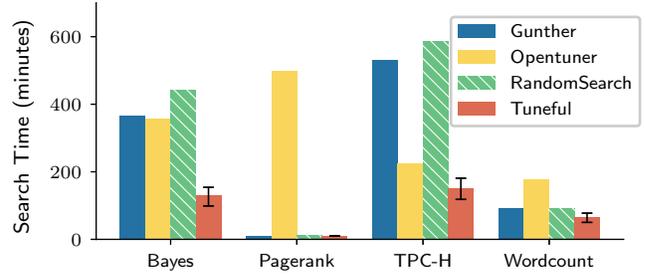}
	\caption{Search time of the different tuning algorithms to find 5\% of the optimal configuration (the lower the better).}
	\label{figure:search_time}
	\vspace*{-15pt}
\end{figure}

For evaluating differences in time taken to \emph{find} configurations close-to-optimal, we allow each algorithm to execute workloads until it finds the first configuration resulting in a runtime within 5\% of the one produced by the \emph{estimated optimal configuration}. This is defined as the best configuration ever found across all our tests for each workload, irrespective of the tuning algorithm or experiment.\\
\noindgras{The median search time for other algorithms is 2.7X longer when compared to Tuneful:}
~\autoref{figure:search_time} shows that for the Bayes workload, OpenTuner, Random Search and Gunther need 2.7X, 3.3X and 2.7X more search time compared to the median Tuneful search time.
For Pagerank, Random Search takes 1.4X the time while Gunther has a comparable performance. However, Opentuner never finds a configuration with a runtime within 5\% of the optimum within the allocated 100 executions. That represents a 49X increase in search time.
For TPC-H, Opentuner, Random Search and Gunther take 2.8X, 3.6X and 3.2X more time than Tuneful, respectively.
For Wordcount, Opentuner takes 2.6X the time while Random Search and Gunther 1.3X.

%

\begin{figure*}[t!]
	\centering
	\vspace*{-30pt}
	\hspace*{-20pt}
	\includegraphics[]{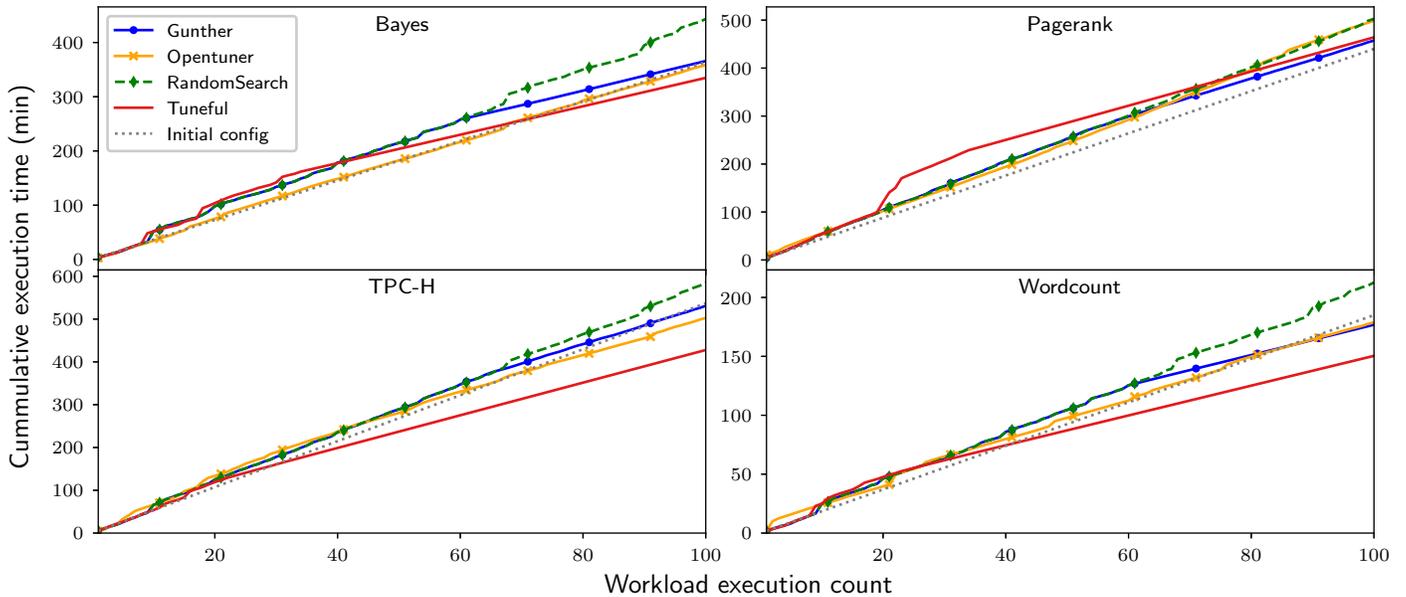}%
	\caption[]{Cumulative execution time over 100 workload executions, under iterative tuning. Shallow slopes represent better configurations (smaller time increment for executing the workload once). Steeper slopes represent worse configurations.}%
	\label{figure:acc_time}%
\end{figure*}

The search time in Tuneful is the sum of the \emph{workload execution times} needed by the tuning algorithm to: (i) explore for significant configuration parameters, (ii) tune those to their optimal values. For each execution, we include the time spent in the tuning algorithm for actions such as sampling, running Bayesian optimisation on GP, etc. We run 2 SA rounds, each using 10 execution samples, followed by tuning (15 execution samples at maximum).
The reported search time does not only depend on the number of samples but also on the actual samples that are picked, as exploring a bad configuration leads to a slow execution of the workload. The GP model suggests samples that most likely have the minimum execution time, leaving others unexplored. However, it still needs to explore the configuration space in order to build an accurate cost model.
Finally, the algorithm overhead is negligible: a few seconds to run and pick the next configuration for exploration/tuning. This is due to the small number of samples that we use during the SA rounds and optimization, in addition to restricting the tuning to the influential parameters. The search time for the other three systems is significantly higher than Tuneful, mostly due to them not being data-efficient in finding close-to-optimal configurations.

We estimate the search cost based on GCP's~\cite{gcp} per-second pricing.
The total cost for tuning the four workloads is  \$379, \$354, \$288 using Opentuner, Gunther and Random Search, respectively. In comparison, tuning the four workloads with Tuneful costs \$94.


\noindgras{Tuneful accelerates the amortization of tuning costs:} the previous experiment does not show the full story on how the different approaches compare in behaviour as they perform incremental tuning from one execution to the next. For that, it is useful to have a timeline view. ~\autoref{figure:acc_time} shows the cumulative execution time of running each workload over multiple configurations, as determined iteratively by the tuning algorithms considered. Here, we start from a plausible developer-guided configuration to reduce exploration costs across the large search space. The dotted line shows cumulative execution time for this config \emph{without any tuning}. 

The tuning "pays off" only after the lines intersect the dotted line, even if good configurations were found much earlier (finding them required exploring some configurations worse than the initial).
Tuneful explores the search space for 35 executions (20 during SA and 15 for tuning), then we pick the best configuration it found and continue only using that. We let other tuning algorithms run longer (100 executions) to see if they find configurations that are better or equivalent to Tuneful's. Better configs are shown as lines with shallower slopes (e.g. when Wordcount is tuned by Gunther, after execution 60), while equivalent configs appear as lines parallel to Tuneful's (e.g. Gunther for the Bayes workload). 

For Pagerank, the initial configuration proved to be a very good one and hard to beat through tuning. While both Tuneful and Gunther find better configurations than it, the exploration cost is not amortized in 100 executions.

%
%
%
%

\begin{figure} [t]
	\centering
	\vspace*{-5mm}
	\hspace*{-5mm}
	\includegraphics[]{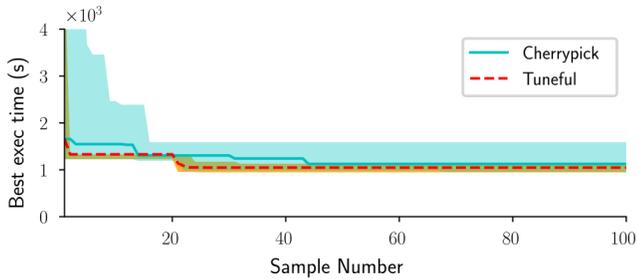}
	\caption{Convergence speed of Cherrypick versus Tuneful (Pagerank workload). The shaded space represents the area between the 90th and 10th percentile of execution time.}
	\label{figure:cherrypick_exp}
	\vspace*{-15pt}
\end{figure}

\noindgras{Tuneful compared to other GP approaches}: By leveraging significance-aware GP optimization, Tuneful is able to find good configurations significantly faster than the other approaches. 
In comparison, it is known that just using GP directly for the high dimensional configuration space will converge slowly or get stuck in local minimums~\cite{tripathy2016gaussian}.
We experimented with Cherrypick~\cite{cherrypick}, which also uses GP to tune the cloud configuration (low dimensional number of parameters). We used it to tune the same 30 configuration parameters for the Pagerank workload and compared against Tuneful. We run the experiments 10 times, each starting with a different initial random configuration, given to both systems (they start exploring the space from the same point).
In~\autoref{figure:cherrypick_exp} the area between the 90th and 10th percentile of execution time is shaded, and the line shows the median of 10 experiments. We plot the minimum execution time found by the tuning algorithm until each sample on the x axis (akin to stopping the tuning at that point and running with the best configuration so far). At the 90th percentile, Cherrypick tries configurations that are more costly (more variation, Cherrypick's median at execution 100 is above the 90th percentile of Tuneful) and takes a significantly longer time to find configurations close to the ones found by Tuneful: Tuneful's median best execution time still 7\% faster than Cherrypick's median at sample 100; Tuneful's 90th percentile is 29\% faster than Cherrypick's.
On the other hand, strategies employing SA on its own, without data-efficient tuning approaches such as GP fail as well: For example, even after reducing the search space, Gunther tunes the 6 significant configuration parameters for the Bayes workload using 2X the search cost needed by Tuneful.


\section{Related Work}
\label{sec:rw}

\noindgras{DISC system Configuration Tuning:} 
Several solutions have been proposed for tuning the configurations of Hadoop/MapReduce workloads.
AROMA~\cite{arome}  is a system for Hadoop resource provisioning and configuration tuning. It uses the k-medoids algorithm to cluster the executed jobs, then leverages Support Vector Machine (SVM) for tuning the configuration.
MROnline\cite{mronline} proposed a modified Hill climbing technique to find good configurations; it limits the search space using predefined tuning rules. 
StarFish~\cite{starfish} is a tuning system inspired by the self-tuning database systems.
It uses a configuration tuning approach in which cost estimation is derived by a What-If engine, predicting the cost of different configurations given some profiled data.
Here, finding good configurations hinges on the accuracy of the what-if engine itself.
Those MapReduce tuning systems fit the Hadoop workloads well, but the number of tuned configuration parameters (6-12 parameters) is significantly smaller than the number of available Spark configuration parameters.

More recently, some work has proposed tuning Spark configurations. 
Yu et al. proposed DAC~\cite{yu2018datasize}, a data-size aware Spark configuration tuning system, using a hierarchical modelling approach to approximate workload execution time as a function of input data-size and configuration. It then leverages Genetic algorithms to search for good configurations based on the execution time estimated by the model.
Tuneful does not incur the high data collection costs of DAC when modelling execution time as a function of a given configuration.
Those high costs are hard to amortize before re-tuning is needed (e.g. because of environment change).
DAC improves performance by 30-80X with respect to the default configuration and tunes 41 configuration parameters in Spark 1.6. Tuneful considers all parameters (30) of a more recent Spark release (some of the parameters have been deprecated).
Wang et al.~\cite{wang2016novel} leverages regression trees to tune Spark configurations; they tune 16 configuration parameters and improve performance by 36\%.
This approach also needs a significant number of execution samples to build an accurate regression tree model.
BestConfig~\cite{bestconfig} is a general-purpose tuning system that uses a divide-and-diverge sampling method and a recursive bound-and-search algorithm to tune configurations.
BestConfig was used to tune 30 spark configuration parameters using 500 execution samples and achieved 80\% runtime performance improvement with respect to the default configuration.
Zhao et al.~\cite{zhao2016adaptive} proposed an Adaptive Serialization Strategy to improve Spark performance. It tunes Spark's serialization strategy dynamically based on runtime statistics.
Tuneful differs from this work as we tune a large set of Spark configuration parameters, not just the serialization strategy.
Chiba et al.~\cite{chiba2016workload} characterizes the memory, network, JVM, and garbage collection usages to tune the performance of Spark,
targeting just the the TPC-H workloads. We evaluate the behavior of Tuneful across a more diverse set of workloads.
Unlike this earlier work, our approach maintains a significantly lower tuning cost, making relatively frequent workload re-tuning possible.

\noindgras{Cloud Configuration (CC) Tuning:} 
Some other systems have looked at configuring cloud instances (number of instances, size and number of cores).
Cherrypick\cite{cherrypick} finds near-optimal cloud configurations, by
leveraging standard GP to build a performance model that allows picking good configurations using a small number of samples. This work suits well the lower dimensional space of cloud configurations, but is harder to apply directly to the higher-dimensionality search spaces of Spark.
PARIS~\cite{paris} is a system for selecting the best VM for certain workloads based on user defined metrics. 
It uses offline profiling for benchmarking various VM types, then combines this with an online fingerprint of each workload. 
The combined data is used to build a decision tree and a random forest-based performance model to select the best VM type. 
Tuneful can work together with these CC tuning systems to enable optimization at the DISC system level after finding the best cloud configuration.

In other domains, relevant work but less cost-efficient has been proposed: Ottertune~\cite{ottertune} and Rafiki~\cite{rafiki} requires hundreds of trials in an offline phase to define a system-wide significant parameters for the DBMS workloads. FLASH~\cite{flash-optimizer} starts with 30-50 execution samples to build decision tree model then incrementally execute more samples till it converges.

BOAT~\cite{dalibard2017boat}, a BO based autotuner that enables developers to provide contextual information, in the form of a bespoke probabilistic model of the workload's behaviour, to accelerate tuning convergence. Tuneful adopts a similar idea by defining workload-specific significant parameters, while performing this incrementally without any developer's intervention.


\section{Conclusion and Future Work}
\label{sec:conclusion}
We presented Tuneful,
an approach for data-efficient configuration tuning through leveraging workload-specific SA and GP.
We showed that Tuneful significantly reduces the exploration costs and accelerates the amortization of tuning costs, while online finding configurations that are comparable to the ones from state-of-the-art tuning algorithms.
We illustrated how Tuneful is designed to be integrated into Spark to provide an efficient configuration tuning with negligible overhead.
As a future work, we will investigate employing workload characterization to accurately define the need for workload re-tuning, and study the applicability of Tuneful in other DISC systems. 



\bibliographystyle{plain}
\bibliography{tuneful}

\end{document}